\documentclass[12pt,epsfig]{article}
\usepackage{amsfonts}
\usepackage{epsfig}
\begin{document}

\vspace{0.2cm}

\begin{center}
{\large\bf Multiple seesaw mechanisms of neutrino masses at the
TeV scale}
\end{center}

\vspace{0.5cm}
\begin{center}
{\bf Zhi-zhong Xing} \footnote{E-mail: xingzz@ihep.ac.cn} ~ and ~
{\bf Shun Zhou} \footnote{E-mail: zhoush@ihep.ac.cn} \\
\vspace{0.2cm} {\small\sl Institute of High Energy Physics and
Theoretical Physics Center for Science Facilities, \\ Chinese
Academy of Sciences, P.O. Box 918, Beijing 100049, China}
\end{center}

\vspace{3cm}

\begin{abstract}
In pursuit of a balance between theoretical naturalness and
experimental testability, we propose two classes of multiple
seesaw mechanisms at the TeV scale to understand the origin of
tiny neutrino masses. They are novel extensions of the canonical
and double seesaw mechanisms, respectively, by introducing even
and odd numbers of gauge-singlet fermions and scalars. It is
thanks to a proper implementation of the global ${\rm U(1)} \times
\mathbb{Z}^{}_{2N}$ symmetry that the overall neutrino mass matrix
in either class has a suggestive nearest-neighbor-interaction
pattern. We briefly discuss possible consequences of these
TeV-scale seesaw scenarios, which can hopefully be explored in the
upcoming Large Hadron Collider and precision neutrino experiments,
and present a simple but instructive example of model building.
\end{abstract}

\begin{center}
{\small\it PACS numbers: 14.60.Pq, 12.15.Ff, 13.66.-a, 98.80.Cq}
\end{center}

\newpage

Solar, atmospheric, reactor and accelerator neutrino oscillation
experiments have jointly provided us with very convincing evidence
that three known neutrinos in the Universe must possess tiny and
non-degenerate rest masses \cite{PDG}. This great breakthrough is
hopefully opening a low-energy window onto new physics beyond the
Standard Model (SM) at very high energy scales. So far many
theoretical and phenomenological attempts have been made towards
understanding the observed neutrino mass hierarchy and lepton flavor
mixing, and among them the {\it seesaw} ideas \cite{SS1,SS2,SS3} are
most brilliant and might even lead us to a true theory of neutrino
masses.

The canonical (type-I) seesaw mechanism \cite{SS1} can naturally
work at a superhigh energy scale $\Lambda^{}_{\rm SS} \sim 10^{14}$
GeV to generate tiny neutrino masses of order $\Lambda^2_{\rm
EW}/\Lambda^{}_{\rm SS} \sim 0.1$ eV with $\Lambda^{}_{\rm EW} \sim
10^2$ GeV being the electroweak scale. To be more specific, the
effective Majorana mass matrix of three light neutrinos is given by
$M^{}_\nu = - M^{}_{\rm D} M^{-1}_{\rm R} M^T_{\rm D}$ in the
leading-order approximation, where $M^{}_{\rm D} \sim {\cal
O}(\Lambda^{}_{\rm EW})$ originates from the Yukawa interactions
between the SM lepton doublet $\ell^{}_{\rm L}$ and the right-handed
neutrinos $N^i_{\rm R}$ (for $i=1,2,3$), and $M^{}_{\rm R} \sim
{\cal O}(\Lambda^{}_{\rm SS})$ is a symmetric matrix coming from the
lepton-number-violating Majorana mass term of $N^{i}_{\rm R}$. This
seesaw picture is technically natural because it allows the relevant
Yukawa couplings to be ${\cal O}(1)$ and requires little fine-tuning
of the textures of $M^{}_{\rm D}$ and $M^{}_{\rm R}$, but it loses
the direct testability on the experimental side and causes a
hierarchy problem on the theoretical side (as long as
$\Lambda^{}_{\rm SS} > 10^7$ GeV \cite{hierarchy}). A possible way
out of the impasse is to lower the seesaw scale down to
$\Lambda^{}_{\rm SS} \sim 1$ TeV, an energy frontier to be soon
explored by the Large Hadron Collider (LHC). However, to test such a
TeV seesaw scenario necessitates an appreciable magnitude of
$M^{}_{\rm D}/M^{}_{\rm R}$ so as to make it possible to produce and
detect heavy Majorana neutrinos at the LHC via their charged-current
interactions. This prerequisite unavoidably requires a terrible
fine-tuning of $M^{}_{\rm D}$ and $M^{}_{\rm R}$, because one has to
impose $M^{}_{\rm R} \sim 1$ TeV, $M^{}_{\rm D}/M^{}_{\rm R} \sim
10^{-3} \cdots 10^{-1}$ and $M^{}_\nu \sim 0.1$ eV simultaneously on
the above seesaw relation \cite{KS}. It is therefore desirable to
invoke new ideas to resolve this {\it unnaturalness} problem built
in the TeV seesaw mechanism.

We stress that a {\it multiple seesaw mechanism} at the TeV scale
may satisfy both {\it naturalness} and {\it testability}
requirements. To illustrate, we assume that the small mass scale of
three light neutrinos arises from a naive seesaw relation $m \sim
(\lambda \Lambda^{}_{\rm EW})^{n+1}/\Lambda^n_{\rm SS}$, where
$\lambda$ is a dimensionless Yukawa coupling coefficient and $n$ is
an arbitrary integer larger than one. Without any terrible
fine-tuning, the seesaw scale can be estimated from
\begin{eqnarray}
\Lambda^{}_{\rm SS} \; \sim \; \lambda^{\frac{n+1}{n}}
\left[\frac{\Lambda^{}_{\rm EW}}{100~{\rm
GeV}}\right]^{\frac{n+1}{n}} \left[\frac{0.1~{\rm eV}}{
m}\right]^{\frac{1}{n}} 10^{\frac{2\left(n+6\right)}{n}}~ {\rm
GeV} \; .
\end{eqnarray}
A numerical change of $\Lambda^{}_{\rm SS}$ with $n$ and $\lambda$
is shown in Fig. 1, where $\Lambda^{}_{\rm SS} \sim 1$ TeV may
naturally result from $n \geq 2$ and $\lambda \geq 10^{-3}$. Hence
the multiple seesaw idea is expected to work at the TeV scale and
provide us with a novel approach to bridge the gap between
theoretical naturalness and experimental testability of the
canonical seesaw mechanism.

The simplest way to build a multiple seesaw model at the TeV scale
is to extend the canonical seesaw mechanism by introducing a
number of gauge-singlet fermions $S^i_{n{\rm R}}$ and scalars
$\Phi^{}_n$ (for $i=1,2,3$ and $n=1,2,\cdots$). We find that a
proper implementation of the global ${\rm U(1)} \times
\mathbb{Z}^{}_{2N}$ symmetry leads us to two classes of multiple
seesaw mechanisms with the nearest-neighbor-interaction pattern
--- an interesting form of the overall
$3\left(n+2\right) \times 3\left(n+2\right)$ neutrino mass matrix
in which every $3\times 3$ sub-matrix only interacts with its
nearest neighbor. The first class contains an even number of
$S^i_{n{\rm R}}$ and $\Phi^{}_n$ and corresponds to an appealing
extension of the canonical seesaw mechanism, while the second
class has an odd number of $S^i_{n{\rm R}}$ and $\Phi^{}_n$ and is
actually a straightforward extension of the double seesaw
mechanism \cite{double}. Their possible collider signatures and
low-energy consequences, together with a simple example of model
building, will be briefly discussed.


The spirit of multiple seesaw mechanisms is to make a harmless
extension of the SM by adding three right-handed neutrinos $N^i_{\rm
R}$ together with some gauge-singlet fermions $S^i_{n{\rm R}}$ and
scalars $\Phi^{}_n$ (for $i=1,2,3$ and $n=1,2,\cdots$). Allowing for
lepton number violation to a certain extent, we can write the
gauge-invariant Lagrangian for neutrino masses as
\begin{eqnarray}
-{\cal L}^{}_\nu = \overline{\ell^{}_{\rm L}} Y^{}_\nu \tilde{H}
N^{}_{\rm R} + \overline{N^c_{\rm R}} Y^{}_{S^{}_1} S^{}_{1{\rm
R}} \Phi^{}_1 + \sum^n_{i = 2} \overline{S^c_{(i-1){\rm R}}}
Y^{}_{S^{}_i} S^{}_{i{\rm R}} \Phi^{}_i + \frac{1}{2}
\overline{S^c_{n{\rm R}}} M^{}_\mu S^{}_{n{\rm R}} + {\rm h.c.} \;
,
\end{eqnarray}
where $\ell^{}_{\rm L}$ and $\tilde{H} \equiv i\sigma^{}_2 H^*$
stand respectively for the $\rm SU(2)^{}_{\rm L}$ lepton and Higgs
doublets, $Y^{}_\nu$ and $Y^{}_{S_i}$ (for $i=1, 2, \cdots, n$)
are the $3\times 3$ Yukawa coupling matrices, and $M^{}_\mu$ is a
symmetric Majorana mass matrix. After spontaneous gauge symmetry
breaking, we arrive at the overall $3\left(n+2\right) \times
3\left(n+2\right)$ neutrino mass matrix $\cal M$ in the flavor
basis defined by $(\nu^{}_{\rm L}, N^c_{\rm R}, S^c_{1{\rm R}},
\cdots, S^c_{n{\rm R}})$ and their charge-conjugate states:
\begin{eqnarray}
{\cal M} = \left(\matrix{{\bf 0} & M^{}_{\rm D} & {\bf 0} & {\bf
0} & {\bf 0} & \cdots & {\bf 0} \cr M^T_{\rm D} & {\bf 0} & M^{}_{
S^{}_1} & {\bf 0} & {\bf 0} & \cdots & {\bf 0} \cr {\bf 0} &
M^T_{S^{}_1} & {\bf 0} & M^{}_{S^{}_2} & {\bf 0} & \cdots & {\bf
0} \cr {\bf 0} & {\bf 0} & M^T_{S^{}_2} & {\bf 0} & \ddots &
\ddots & \vdots \cr {\bf 0} & {\bf 0} & {\bf 0} & \ddots & \ddots
& M^{}_{S^{}_{n-1}} & {\bf 0} \cr \vdots & \vdots & \vdots &
\ddots & M^T_{S^{}_{n-1}} & {\bf 0} & M^{}_{S^{}_n} \cr {\bf 0} &
{\bf 0} & {\bf 0} & \cdots & {\bf 0} & M^T_{S^{}_n} &
M^{}_\mu}\right) \; ,
\end{eqnarray}
where $M^{}_{\rm D} \equiv Y^{}_\nu \langle H \rangle$ and
$M^{}_{S^{}_i} = Y^{}_{S^{}_i} \langle \Phi^{}_i \rangle$ (for $i
= 1 ,2, \cdots, n$) are $3\times 3$ mass matrices. Setting
$N^{}_{\rm R} = S^{}_{0{\rm R}}$ for simplicity, one can observe
that the Yukawa interactions between $S^{}_{i{\rm R}}$ and
$S^{}_{j{\rm R}}$ exist if and only if their subscripts satisfy
the selection rule $|i - j| =1$ (for $i,j = 0, 1, 2, \cdots, n$).
Note that $\cal M$ manifests a very suggestive
nearest-neighbor-interaction pattern, which has attracted a lot of
attention in the quark sector to understand the observed
hierarchies of quark masses and flavor mixing angles
\cite{Fritzsch}. Such a special structure of $\cal M$, or
equivalently that of ${\cal L}^{}_\nu$ in Eq. (2), may arise from
a proper implementation of the global ${\rm U(1)}\times
\mathbb{Z}^{}_{2N}$ symmetry. The unique generator of the cyclic
group $\mathbb{Z}^{}_{2N}$ is $\varpi = e^{i\pi/N}$, which
produces all the group elements $\mathbb{Z}^{}_{2N} = \{1, \varpi,
\varpi^2, \varpi^3, \cdots, \varpi^{2N-1}\}$. By definition, a
field $\Psi$ with the charge $q$ transforms as $\Psi \to e^{i\pi
q/N} \Psi$ under $\mathbb{Z}_{2N}$ (for $q = 0, 1, 2, \cdots,
2N-1$). Hence we manage to assign the $\rm U(1)$ and
$\mathbb{Z}^{}_{2N}$ charges of the relevant fields in Eq. (2) in
the following way:
\begin{enumerate}
\item The global $\rm U(1)$ symmetry can be identified with the
lepton number $L$, namely $L(\ell^{}_{\rm L}) = L(E^{}_{\rm R}) = +
1$, where $E^{}_{\rm R}$ represents the charged-lepton singlets in
the SM. We arrange the lepton numbers of gauge-singlet fermions and
scalars to be $L(N^{}_{\rm R}) = +1$, $L(S^{}_{k{\rm R}}) = (-1)^k$
and $L(\Phi^{}_k) = 0$ (for $k=1, 2, \cdots, n$). It turns out that
only the Majorana mass term $\overline{S^c_{n{\rm R}}} M^{}_\mu
S^{}_{n{\rm R}}$ in ${\cal L}^{}_\nu$ explicitly violates the $\rm
U(1)$ symmetry. After this assignment, other lepton-number-violating
mass terms (e.g., $\overline{N^c_{\rm R}} M^{}_{\rm R} N^{}_{\rm R}$
in the canonical seesaw mechanism) may also appear in the
Lagrangian, but they can be eliminated by invoking the discrete
$\mathbb{Z}^{}_{2N}$ symmetry.

\item We assign the $\mathbb{Z}^{}_{2N}$ charge of $S^{}_{n{\rm
R}}$ as $q(S^{}_{n{\rm R}}) = N$. Then it is easy to verify that
the Majorana mass term $\overline{S^c_{n{\rm R}}} M^{}_\mu
S^{}_{n{\rm R}}$ is invariant under the $\mathbb{Z}^{}_{2N}$
transformation. If all the other gauge-singlet fermions
$S^{}_{k{\rm R}}$ (for $k = 1, 2, \cdots, n-1$) take any charges
in $\{1, 2, \cdots, 2N-1\}$ other than $N$, their corresponding
Majorana mass terms are accordingly forbidden. Given
$q(\ell^{}_{\rm L}) = q(E^{}_{\rm R}) = q(N^{}_{\rm R}) = 1$, both
the charges of $S^{}_{k{\rm R}}$ (for $k = 1, 2, \cdots, n-1$) and
those of $\Phi^{}_i$ (for $i = 1, 2, \cdots, n$) can be properly
chosen so as to achieve the nearest-neighbor-interaction form of
${\cal L}^{}_\nu$ as shown in Eq. (2). But the solution to this
kind of charge assignment may not be unique, because for a given
value of $n$ one can always take $N \gg n$ to fulfill all the
above-mentioned requirements \cite{XZ09}. Simple examples (with $n
= 1, 2, 3$) will be presented below.
\end{enumerate}
We remark that our multiple seesaw picture should be the simplest
extension of the canonical seesaw mechanism, since it does not
invoke the help of either additional ${\rm SU(2)}^{}_{\rm L}$
fermion doublets \cite{Dudas} or a new isospin 3/2 Higgs multiplet
\cite{Babu}. We also stress that the double seesaw scenario
\cite{double} is only the simplest example in one class of our
multiple seesaw mechanisms (with an odd number of $S^i_{n{\rm R}}$
or $\Phi^{}_n$) and cannot reflect any salient features of the
other class of multiple seesaw mechanisms (with an even number of
$S^i_{n{\rm R}}$ or $\Phi^{}_n$).

Now let us diagonalize $\cal M$ in Eq. (3) to achieve the
effective mass matrix of three light neutrinos $M^{}_\nu$ in the
multiple seesaw mechanisms. Note that $\cal M$ can be rewritten as
\begin{eqnarray}
{\cal M} = \left(\matrix{{\bf 0} & \tilde{M}^{}_{\rm D} \cr
\tilde{M}^T_{\rm D} & \tilde{M}^{}_\mu}\right) \; ,
\end{eqnarray}
where $\tilde{M}^{}_{\rm D} = \left(M^{}_{\rm D} ~~ {\bf 0}\right)$
denotes a $3\times 3\left(n+1 \right)$ matrix and $\tilde{M}^{}_\mu$
is a symmetric $3\left(n+1 \right)\times 3\left(n+1 \right)$ matrix.
Taking the mass scale of $\tilde{M}^{}_\mu$ to be much higher than
that of $\tilde{M}^{}_{\rm D}$, one can easily obtain $M^{}_\nu = -
\tilde{M}^{}_{\rm D} \tilde{M}^{-1}_\mu \tilde{M}^T_{\rm D}$ for
three light Majorana neutrinos in the leading-order approximation.
Because the elements in the fourth to $3n$-th columns of
$\tilde{M}^{}_{\rm D}$ are exactly zero, only the $3\times 3$ top
left block of $\tilde{M}^{-1}_\mu$ is relevant to the calculation of
$M^{}_\nu$. Without loss of generality, the inverse of
$\tilde{M}^{}_\mu$ can be figured out by assuming all the non-zero
$3\times 3$ sub-matrices of ${\cal M}$ to be of rank three. We find
two types of solutions \cite{XZ09}, depending on whether $n$ is even
or odd, and thus arrive at two classes of multiple seesaw
mechanisms:

{\it Class A of multiple seesaw mechanisms} --- they contain an
even number of gauge-singlet fermions $S^i_{n{\rm R}}$ and scalars
$\Phi^{}_n$ (i.e., $n=2k$ with $k=0,1,2,\cdots$) and correspond to
a novel extension of the canonical seesaw picture. The effective
mass matrix of three light Majorana neutrinos is given by
\begin{eqnarray}
M^{}_\nu = - M^{}_{\rm D} \left[\prod^{k}_{i=1}
\left(M^T_{S^{}_{2i-1}}\right)^{-1} M^{}_{S^{}_{2i}} \right]
M^{-1}_\mu \left[\prod^{k}_{i=1}
\left(M^T_{S^{}_{2i-1}}\right)^{-1} M^{}_{S^{}_{2i}} \right]^T
M^T_{\rm D} \;
\end{eqnarray}
in the leading-order approximation. The $k=0$ case is obviously
equivalent to the canonical seesaw mechanism (i.e., $M^{}_\nu = -
M^{}_{\rm D} M^{-1}_{\rm R} M^T_{\rm D}$ by setting $S^{}_{0\rm R} =
N^{}_{\rm R}$ and $M^{}_\mu = M^{}_{\rm R}$). If $M^{}_{S_{2i}} \sim
M^{}_{\rm D} \sim {\cal O}(\Lambda^{}_{\rm EW})$ and
$M^{}_{S_{2i-1}} \sim M^{}_\mu \sim {\cal O}(\Lambda^{}_{\rm SS})$
hold (for $i = 1, 2, \cdots, k$), Eq. (5) leads to $M^{}_\nu \sim
\Lambda^{2\left(k+1\right)}_{\rm EW}/\Lambda^{2k+1}_{\rm SS}$, which
can effectively lower the conventional seesaw scale $\Lambda^{}_{\rm
SS} \sim 10^{14}$ GeV down to the TeV scale as illustrated in Fig.
1.

Taking $n = 2$ (i.e., $k=1$) for example \cite{Gavela}, we arrive
at the minimal extension of the canonical seesaw mechanism:
\begin{eqnarray}
M^{}_\nu = - M^{}_{\rm D} \left(M^T_{S^{}_1}\right)^{-1}
M^{}_{S^{}_2} M^{-1}_\mu M^T_{S^{}_2}
\left(M^{}_{S^{}_1}\right)^{-1} M^T_{\rm D} \; .
\end{eqnarray}
This effective multiple seesaw mass term is illustrated in Fig.
2(a). The nearest-neighbor-interaction pattern of ${\cal M}$ with
$n=2$ can be obtained by imposing the global ${\rm U(1)} \times
\mathbb{Z}^{}_{6}$ symmetry on ${\cal L}^{}_\nu$, in which the
proper charge assignment is listed in Table 1.
\begin{table}[t]
\begin{center}
\caption{The charges of relevant fermion and scalar fields under the
$\rm U(1) \times \mathbb{Z}^{}_6$ symmetry in the multiple seesaw
mechanism with $n=2$.} \vspace{0.3cm}
\begin{tabular}{c|c|c|c|c|c|c|c|c}
  \hline
  \hline
  $~$ & $\ell^{}_{\rm L}$ & $H$ & $E^{}_{\rm R}$ & $N^{}_{\rm R}$ &
  $S^{}_{1{\rm R}}$ & $S^{}_{2{\rm R}}$ & $\Phi^{}_1$ & $\Phi^{}_2$  \\
  \hline
  $L$ & $+1$ & $0$ & $+1$ & $+1$ & $-1$ & $+1$ & $0$ & $0$\\
  \hline
  $q$ & $+1$ & $0$ & $+1$ & $+1$ & $+2$ & $+3$ & $+3$ & $+1$\\
  \hline
\end{tabular}
\end{center}
\end{table}

{\it Class B of multiple seesaw mechanisms} --- they contain an
odd number of gauge-singlet fermions $S^i_{n{\rm R}}$ and scalars
$\Phi^{}_n$ (i.e., $n=2k+1$ with $k=0,1,2,\cdots$) and correspond
to an interesting extension of the double seesaw picture. The
effective mass matrix of three light Majorana neutrinos reads
\small
\begin{eqnarray}
M^{}_\nu = M^{}_{\rm D} \left[\prod^{k}_{i=1}
\left(M^T_{S^{}_{2i-1}}\right)^{-1} M^{}_{S^{}_{2i}} \right]
\left(M^T_{S^{}_{2k+1}}\right)^{-1} M^{}_\mu \left(M^{}_{
S^{}_{2k+1}}\right)^{-1} \left[\prod^{k}_{i=1}
\left(M^T_{S^{}_{2i-1}}\right)^{-1} M^{}_{S^{}_{2i}} \right]^T
\hspace{-0.15cm} M^T_{\rm D} \;
\end{eqnarray}
\normalsize in the leading-order approximation. The $k=0$ case just
corresponds to the double seesaw scenario with a very low mass scale
of $M^{}_\mu$ \cite{double}: $M^{}_\nu = M^{}_{\rm D}
\left(M^T_{S^{}_1}\right)^{-1} M^{}_\mu
\left(M^{}_{S^{}_1}\right)^{-1} M^T_{\rm D}$. Note that the
nearest-neighbor-interaction pattern of $\cal M$ in the double
seesaw mechanism is guaranteed by an implementation of the global
${\rm U(1)} \times \mathbb{Z}^{}_4$ symmetry with the following
charge assignment: $L(\ell^{}_{\rm L}) = L(E^{}_{\rm R}) =
L(N^{}_{\rm R}) = +1$, $L(S^{}_{1{\rm R}}) =-1$, $L(H) =
L(\Phi^{}_1) =0$, $q(\ell^{}_{\rm L}) = q(E^{}_{\rm R}) =
q(N^{}_{\rm R}) = q(\Phi^{}_1) =+1$, $q(H) = 0$ and $q(S^{}_{1\rm
R}) =+2$.

If $M^{}_{S_{2i}} \sim M^{}_{\rm D} \sim {\cal O}(\Lambda^{}_{\rm
EW})$ and $M^{}_{S_{2i-1}} \sim {\cal O}(\Lambda^{}_{\rm SS})$
hold (for $i = 1, 2, \cdots, k$), the mass scale of $M^{}_\mu$ is
in general unnecessary to be as small as that given by the double
seesaw mechanism. To be more specific, let us consider the minimal
extension of the double seesaw picture by taking $n=3$. In this
case, we impose the ${\rm U(1)} \times \mathbb{Z}^{}_{10}$
symmetry on ${\cal L}^{}_\nu$ with a proper charge assignment
listed in Table 2 to assure the nearest-neighbor-interaction form
of $\cal M$. The corresponding formula of $M^{}_\nu$ is
\begin{eqnarray}
M^{}_\nu = M^{}_{\rm D} \left(M^T_{S^{}_1}\right)^{-1}
M^{}_{S^{}_2} \left(M^T_{S^{}_3}\right)^{-1} M^{}_\mu
\left(M^{}_{S^{}_3}\right)^{-1} M^T_{S^{}_2}
\left(M^{}_{S^{}_1}\right)^{-1} M^T_{\rm D} \; .
\end{eqnarray}
This effective multiple seesaw mass term is illustrated in Fig.
2(b). It becomes obvious that the proportionality of $M^{}_\nu$ to
$M^{}_\mu$ in Eq. (8) is doubly suppressed not only by the ratio
$M^{}_{\rm D}/M^{}_{S^{}_1} \sim \Lambda^{}_{\rm EW}/\Lambda^{}_{\rm
SS}$ but also by the ratio $M^{}_{S^{}_2}/M^{}_{S^{}_3} \sim
\Lambda^{}_{\rm EW}/\Lambda^{}_{\rm SS}$, and thus $M^{}_\nu \sim
0.1$ eV can naturally result from $Y^{}_\nu \sim Y^{}_{S^{}_1} \sim
Y^{}_{S^{}_2} \sim Y^{}_{S^{}_3} \sim {\cal O}(1)$ and $M^{}_\mu
\sim 1$ keV at $\Lambda^{}_{\rm SS} \sim 1$ TeV.
\begin{table}[t]
\begin{center}
\caption{The charges of relevant fermion and scalar fields under the
$\rm U(1) \times \mathbb{Z}^{}_{10}$ symmetry in the multiple seesaw
mechanism with $n=3$.} \vspace{0.3cm}
\begin{tabular}{c|c|c|c|c|c|c|c|c|c|c}
  \hline
  \hline
  $~$ & $\ell^{}_{\rm L}$ & $H$ & $E^{}_{\rm R}$ & $N^{}_{\rm R}$ &
  $S^{}_{1{\rm R}}$ & $S^{}_{2{\rm R}}$ & $S^{}_{3{\rm R}}$ & $\Phi^{}_1$
  & $\Phi^{}_2$ & $\Phi^{}_3$ \\
  \hline
  $L$ & $+1$ & $0$ & $+1$ & $+1$ & $-1$ & $+1$ & $-1$ & $0$ & $0$ & $0$\\
  \hline
  $q$ & $+1$ & $0$ & $+1$ & $+1$ & $+2$ & $+3$ & $+5$ & $+7$ & $+5$ & $+2$\\
  \hline
\end{tabular}
\end{center}
\end{table}
\vspace{-0.55cm}

{\it Charged-current interactions of neutrinos} --- they are
important for both production and detection of light and heavy
Majorana neutrinos in a realistic experiment. To define the
neutrino mass eigenstates, we diagonalize the overall mass matrix
${\cal M}$ in Eq. (4) by means of the following unitary
transformation:
\begin{eqnarray}
\left(\matrix{V & \tilde{R} \cr \tilde{S} &
\tilde{U}}\right)^\dagger \left(\matrix{{\bf 0} & \tilde{M}^{}_{\rm
D} \cr \tilde{M}^T_{\rm D} & \tilde{M}^{}_\mu}\right)
\left(\matrix{V & \tilde{R} \cr \tilde{S} & \tilde{U}}\right)^* =
\left(\matrix{\widehat{M}^{}_\nu & {\bf 0} \cr {\bf 0} &
\widehat{M}^{}_{N+S}}\right) \; ,
\end{eqnarray}
where $\widehat{M}^{}_\nu \equiv {\rm Diag}\{m^{}_1, m^{}_2,
m^{}_3\}$ contains the masses of three light Majorana neutrinos
($\hat{\nu}^{}_1$, $\hat{\nu}^{}_2$, $\hat{\nu}^{}_3$), and
$\widehat{M}^{}_{N+S}$ denotes a diagonal matrix whose eigenvalues
are the masses of $3\left(n+1\right)$ heavy Majorana neutrinos
($\hat{N}$, $\hat{S}^{}_1$, $\cdots$, $\hat{S}^{}_n$; and each of
them consists of three components). The SM charged-current
interactions of $\nu^{}_e$, $\nu^{}_\mu$ and $\nu^{}_\tau$ can
therefore be expressed, in terms of the mass eigenstates of light
and heavy Majorana neutrinos, as
\begin{eqnarray}
-{\cal L}^{}_{\rm cc} = \frac{g}{\sqrt{2}} \overline{\left(\matrix{e
& \mu & \tau}\right)^{}_{\rm L}} ~\gamma^\mu \left[V
\left(\matrix{\hat{\nu}^{}_1 \cr \hat{\nu}^{}_2 \cr
\hat{\nu}^{}_3}\right)^{}_{\rm L} + \tilde{R} \left(\matrix{\hat{N}
\cr \hat{S}^{}_1 \cr \vdots \cr \hat{S}^{}_n}\right)^{}_{\rm L}
\right] W^-_\mu + {\rm h.c.} \;
\end{eqnarray}
in the basis where the mass eigenstates of three charged leptons
are identified with their flavor eigenstates. Note that $V$ is the
$3\times 3$ neutrino mixing matrix responsible for neutrino
oscillations, while the $3\times 3\left(n+1\right)$ matrix
$\tilde{R}$ governs the strength of charged-current interactions
of heavy Majorana neutrinos. Note also that both $V V^\dagger +
\tilde{R} \tilde{R}^\dagger = {\bf 1}$ and $V\widehat{M}^{}_\nu
V^T + \tilde{R} \widehat{M}^{}_{N+S} \tilde{R}^T = {\bf 0}$ hold,
and thus $V$ must be non-unitary. It is $\tilde{R}$ that measures
the deviation of $V$ from unitarity in neutrino oscillations and
determines the collider signatures of heavy Majorana neutrinos at
the LHC.


We expect that our multiple seesaw idea can lead to rich
phenomenology at both the TeV scale and lower energies. For
simplicity, here we only mention a few aspects of the
phenomenological consequences of multiple seesaw mechanisms.
\begin{itemize}
\item     {\it Non-unitary neutrino mixing and CP violation}.
Since $V$ is non-unitary, it generally involves  a number of new
flavor mixing parameters and new CP-violating phases \cite{X08}.
Novel CP-violating effects in the medium-baseline $\nu^{}_\mu \to
\nu^{}_\tau$ and $\overline{\nu}^{}_\mu \to \overline{\nu}^{}_\tau$
oscillations may therefore show up and provide a promising signature
of the unitarity violation of $V$, which could be measured at a
neutrino factory \cite{CP}.

\item     {\it Signatures of heavy Majorana neutrinos at the LHC}.
Given $\tilde{M}^{}_{\rm D} \sim {\cal O}(\Lambda^{}_{\rm EW})$ and
$\tilde{M}^{}_\mu \sim {\cal O}(\Lambda^{}_{\rm SS}) \sim {\cal
O}(1)$ TeV, it is straightforward to obtain $\tilde{R} \approx
\tilde{M}^{}_{\rm D} \tilde{M}^{-1}_\mu \tilde{U} \sim {\cal
O}(0.1)$, which actually saturates the present experimental upper
bound on $|\tilde{R}|$ \cite{Antusch}. For {\it Class A} of multiple
seesaw mechanisms, their clear LHC signatures are expected to be the
like-sign dilepton events arising from the lepton-number-violating
processes $pp \to l^\pm_\alpha l^\pm_\beta X$ (for $\alpha, \beta =
e, \mu, \tau$) mediated by heavy Majorana neutrinos \cite{Han}. For
{\it Class B} of multiple seesaw mechanisms with $M^{}_\mu \ll
\Lambda^{}_{\rm EW}$, the mass spectrum of heavy Majorana neutrinos
generally exhibits a pairing phenomenon in which the
nearest-neighbor Majorana neutrinos have nearly degenerate masses
and can be combined to form pseudo-Dirac particles. This feature has
already been observed in the double seesaw model \cite{double}.
Therefore, the discriminating collider signatures at the LHC are
expected to be the $pp \to l^\pm_\alpha l^\pm_\beta l^\mp_\gamma X$
processes (for $\alpha, \beta, \gamma = e, \mu, \tau$)
\cite{triple}.

\item     {\it Possible candidates for dark matter}. One or more of
the heavy Majorana neutrinos and gauge-singlet scalars in our
multiple seesaw mechanisms could be arranged to have a
sufficiently long lifetime. Such weakly-interacting and massive
particles might therefore be a plausible candidate for cold dark
matter \cite{DM}.
\end{itemize}
One may explore more low-energy effects of multiple seesaw
mechanisms, such as their contributions to the
lepton-flavor-violating processes $\mu \rightarrow e\gamma$ and so
on. It should also be interesting to explore possible baryogenesis
via leptogenesis \cite{FY}, based on a multiple seesaw picture, to
interpret the cosmological matter-antimatter asymmetry.

As a flexible and testable TeV seesaw scheme, the multiple seesaw
mechanisms can also provide us with plenty of room for model
building. But the latter requires further inputs or assumptions.
Here we present a simple but instructive example, in which all the
textures of $3\times 3$ sub-matrices in the overall neutrino mass
matrix ${\cal M}$ are symmetric and have the well-known Fritzsch
pattern \cite{Fritzsch},
\begin{eqnarray}
M^{}_a \; =\; \left( \matrix{ 0 & ~ x^{}_{a} ~ & 0 \cr x^{}_{a} & 0
& y^{}_{a} \cr 0 & y^{}_{a} & z^{}_{a} \cr} \right) \;
\end{eqnarray}
with $a = {\rm D}, S^{}_1, \cdots, S^{}_n$ or $\mu$, for
illustration. Choosing the Fritzsch texture makes sense because it
coincides with the nearest-neighbor-interaction form of $\cal M$
itself. We make an additional assumption that the ratio
$x^{}_a/y^{}_a$ is a constant independent of the subscript $a$.
Then it is easy to show that the effective mass matrix of three
light Majorana neutrinos $M^{}_\nu$ has the same Fritzsch texture
in the leading-order approximation:
\begin{eqnarray}
M^{}_\nu \; =\; -\left( \matrix{ 0 & \displaystyle\frac{x^2_{\rm
D}}{x^{}_\mu} \left[\prod^k_{i=1} \frac{x^2_{S^{}_{2i}}}{x^2_{
S^{}_{2i-1}}}\right] & 0 \cr\cr \displaystyle\frac{x^2_{\rm
D}}{x^{}_\mu} \left[\prod^k_{i=1} \frac{x^2_{S^{}_{2i}}}{x^2_{
S^{}_{2i-1}}}\right] & 0 & \displaystyle\frac{y^2_{\rm D}}{y^{}_\mu}
\left[\prod^k_{i=1} \frac{y^2_{S^{}_{2i}}}{y^2_{
S^{}_{2i-1}}}\right] \cr\cr 0 & \displaystyle\frac{y^2_{\rm
D}}{y^{}_\mu} \left[\prod^k_{i=1} \frac{y^2_{S^{}_{2i}}}{y^2_{
S^{}_{2i-1}}}\right] & \displaystyle\frac{z^2_{\rm D}}{z^{}_\mu}
\left[\prod^k_{i=1} \frac{z^2_{S^{}_{2i}}}{z^2_{
S^{}_{2i-1}}}\right] \cr} \right) \;
\end{eqnarray}
derived from Eq. (5) for {\it Class A of multiple seesaw mechanisms}
(with $n=2k$ for $k=0,1,2,\cdots$); and
\begin{eqnarray}
M^{}_\nu \; =\; \left( \matrix{ 0 & \displaystyle\frac{x^2_{\rm
D}}{x^2_{S^{}_{2k+1}}} \left[\prod^k_{i=1}
\frac{x^2_{S^{}_{2i}}}{x^2_{ S^{}_{2i-1}}}\right] x^{}_\mu & 0
\cr\cr \displaystyle\frac{x^2_{\rm D}}{x^2_{S^{}_{2k+1}}}
\left[\prod^k_{i=1} \frac{x^2_{S^{}_{2i}}}{x^2_{
S^{}_{2i-1}}}\right] x^{}_\mu & 0 & \displaystyle\frac{y^2_{\rm
D}}{y^2_{S^{}_{2k+1}}} \left[\prod^k_{i=1}
\frac{y^2_{S^{}_{2i}}}{y^2_{ S^{}_{2i-1}}}\right] y^{}_\mu \cr\cr 0
& \displaystyle\frac{y^2_{\rm D}}{y^2_{S^{}_{2k+1}}}
\left[\prod^k_{i=1} \frac{y^2_{S^{}_{2i}}}{y^2_{
S^{}_{2i-1}}}\right] y^{}_\mu & \displaystyle\frac{z^2_{\rm
D}}{z^2_{S^{}_{2k+1}}} \left[\prod^k_{i=1}
\frac{z^2_{S^{}_{2i}}}{z^2_{ S^{}_{2i-1}}}\right] z^{}_\mu \cr}
\right) \;
\end{eqnarray}
\normalsize obtained from Eq. (7) for {\it Class B of multiple
seesaw mechanisms} (with $n=2k+1$ for $k=0,1,2,\cdots$). This
seesaw-invariant property of $M^{}_\nu$ is interesting since it
exactly reflects how two classes of multiple seesaw mechanisms
work for every element of $M^{}_\nu$. Note that it is possible to
interpret current experimental data on small neutrino masses and
large flavor mixing angles by taking both the texture of the light
neutrino mass matrix $M^{}_\nu$ and that of the charged-lepton
mass matrix $M^{}_l$ to be of the Fritzsch form \cite{Xing02}.
Hence the above examples are phenomenologically viable. Once the
texture of $M^{}_\nu$ is fully reconstructed from more accurate
neutrino oscillation data, one may then consider to quantitatively
explore the textures of those $3\times 3$ sub-matrices of $\cal M$
in such a multiple seesaw model.


To conclude, new ideas are eagerly wanted in the upcoming LHC era
to achieve a proper balance between theoretical naturalness and
experimental testability of the elegant seesaw pictures, which
ascribe the small masses of three known neutrinos to the existence
of some heavy degrees of freedom. In the present work we have
extended the canonical and double seesaw scenarios and proposed
two classes of multiple seesaw mechanisms at the TeV scale by
introducing an arbitrary number of gauge-singlet fermions and
scalars into the SM and by implementing the global ${\rm U(1)}
\times \mathbb{Z}^{}_{2N}$ symmetry in the neutrino sector. These
new TeV-scale seesaw mechanisms are expected to lead to rich
phenomenology at low energies and open some new prospects for
understanding the origin of tiny neutrino masses and lepton number
violation.

\vspace{0.4cm}

This research was supported in part by the National Natural
Science Foundation of China under grant No. 10425522 and No.
10875131.

\newpage

\begin{figure}[t]
\vspace{-6cm}
\epsfig{file=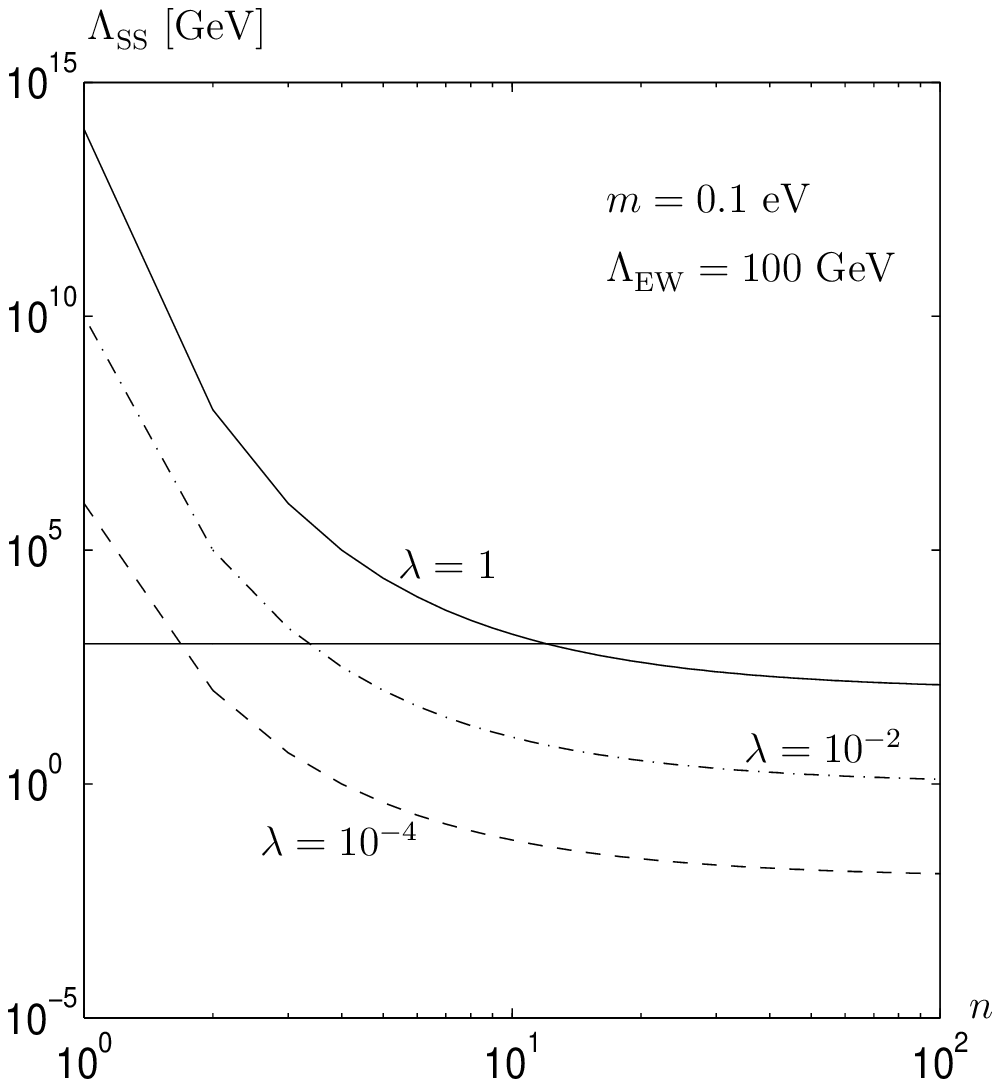,bbllx=-1.0cm,bblly=14.5cm,bburx=9.0cm,bbury=31.5cm,%
width=7cm,height=12cm,angle=0,clip=0} \vspace{3.2cm} \caption{A
numerical illustration of the seesaw scale $\Lambda^{}_{\rm SS}$
changing with $n$ and $\lambda$ as specified in Eq. (1). Here the
horizontal line stands for the TeV scale.}
\end{figure}

\begin{figure}[t]
\vspace{-5cm}
\epsfig{file=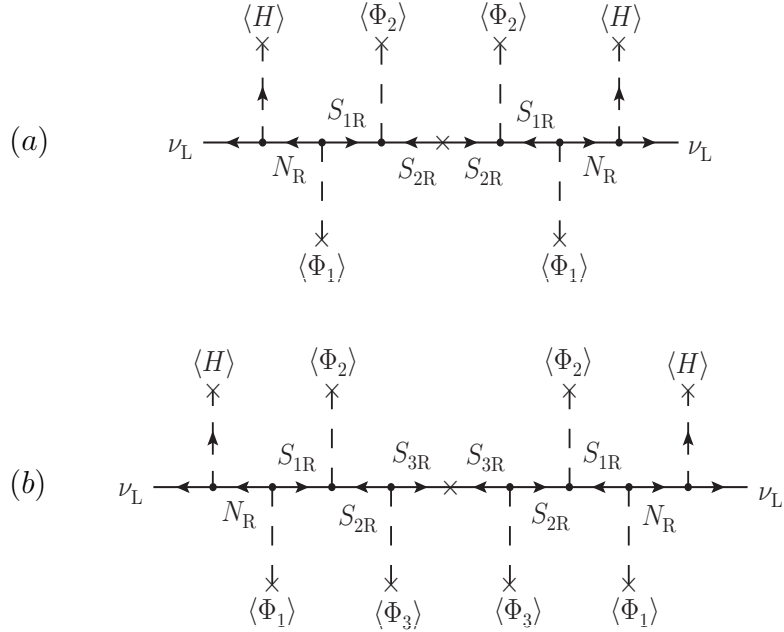,bbllx=2.0cm,bblly=14cm,bburx=12.0cm,bbury=31cm,%
width=10cm,height=17cm,angle=0,clip=0} %
\vspace{-2.1cm} \caption{The origin of light Majorana neutrino
masses in multiple seesaw mechanisms: (a) the minimal extension of
the canonical seesaw mechanism (with $n=2$); and (b) the minimal
extension of the double seesaw mechanism (with $n=3$).}
\end{figure}

\end{document}